\documentclass[conference,letterpaper]{IEEEtran}

\usepackage[utf8]{inputenc} 
\usepackage[T1]{fontenc}
\usepackage{url}
\usepackage{ifthen}
\usepackage{cite}
\usepackage[cmex10]{amsmath} 
														
\usepackage{array}
\usepackage{multirow}
\usepackage{amsthm}
\usepackage{amssymb}
\usepackage{graphicx}
\usepackage{lipsum}
\PassOptionsToPackage{normalem}{ulem}
\usepackage{ulem}
\usepackage{makecell}
\usepackage{algorithm}
\usepackage{algorithmicx}
\usepackage[noend]{algpseudocode}

\providecommand{\tabularnewline}{\\}

\theoremstyle{plain}
\newtheorem{thm}{\protect\theoremname}
\theoremstyle{definition}
\newtheorem{example}[thm]{\protect\examplename}

\usepackage[caption=false,font=footnotesize]{subfig}
\IEEEoverridecommandlockouts

\makeatother

\usepackage[american]{babel}
\providecommand{\examplename}{Example}
\providecommand{\theoremname}{Theorem}

\begin{document}
\title{Pattern-Based Analysis of Time Series: Estimation} 


\author{%
  \IEEEauthorblockN{Elyas Sabeti\thanks{This work was supported in part by Michigan Institute for Data Science and the Prometheus  Program  of the  Defense  Advanced  Research  Projects  Agency  (DARPA),  grant  number N66001-17-2-401.}}
  \IEEEauthorblockA{Michigan Institute for Data Science\\
                    University of Michigan\\
                    Ann Arbor, MI 48109\\
                    Email: sabeti@umich.edu}
  \and
  \IEEEauthorblockN{Peter X.K. Song}
  \IEEEauthorblockA{School of Public Health\\
	Department of Biostatistics\\
                    University of Michigan\\ 
                    Ann Arbor, MI 48109\\
                    Email: pxsong@umich.edu}
										
	\and
  \IEEEauthorblockN{Alfred O. Hero}
  \IEEEauthorblockA{Department of Electrical Engineering\\
	and Computer Science\\
                    University of Michigan\\ 
                    Ann Arbor, MI 48109\\
                    Email: hero@umich.edu}
}


\maketitle

\begin{abstract}
   While Internet of Things (IoT) devices and sensors create continuous
streams of information, Big Data infrastructures are deemed to
handle the influx of data in real-time. One type of such a continuous
stream of information is time series data. Due to the richness of
information in time series and inadequacy of summary statistics to
encapsulate structures and patterns in such data, development of new
approaches to \emph{learn} time series is of interest. In this paper,
we propose a novel method, called pattern tree, to learn patterns
in the times-series using a binary-structured tree. While a pattern
tree can be used for many purposes such as lossless compression, prediction
and anomaly detection, in this paper we focus on its application in time series estimation and forecasting. In comparison to other methods, our proposed pattern tree method improves the mean squared error of estimation.
\vspace{-0.05in}
\end{abstract}


\section{Introduction}
\vspace{-0.05in}
With the enormous amount of data being collected by the \emph{smart}
devices and sensors, Internet of Things (IoT), high resolution time series
are becoming one of the main types of Big Data. Due to the long temporal
length and high temporal resolution of time series, efficient processing
of such data can be challenging. One possibility is characterization
through low order statistics, but such approaches may lead to the loss of information-rich
parts of time series, since in many applications most of the value
in the information is in the parts that deviates from the low order statistics,
i.e. in the atypical parts \cite{sabeti2019data,host2019data}. For
example, consider the heart rate time series recorded by wearables
that are being used as remote health monitoring devices (IoT in health
care \cite{islam2015internet}), there are short-duration arrhythmic
patterns that are valuable in the sense that they are known to indicate
possible onset of disease, but not fully reflected in simple summary statistics like the sample mean, variance or covariance. Another
example is stock market data, for which hourly stock data is more
eventful than the daily data. Hence, the availability of such time series
data and the richness of information they contain encourage new approaches
in order to efficiently \emph{learn} the structure of the time series, and extract patterns for more accurate processing (e.g.
estimation, prediction and detection).

Regression and classification trees (CART) and random forests \cite{breiman2017classification}  constitute a general approach to non-parametric analysis of data developed in the early 1980's. More recently, model-free pattern-based analysis has been considered in the
literature \cite{chung2004evolutionary,ouyang2010ordinal,liu2011novel,berndt1994using,fu2007stock,alvisi2007short,alvarez2010energy,teng2003regression,hu2014pattern}. Kozat et al. \cite{kozat2007universal} used a universal prediction approach, called a context tree to partition the regressors space
resulting in a piece-wise linear prediction technique.
Chung et al. \cite{chung2004evolutionary} developed a dynamic approach
using evolutionary computation for pattern-based segmentation of time series.
Ouyang et al. \cite{ouyang2010ordinal} proposed a dissimilarity measure
based on ordinal patterns of EEG (electroencephalogram) time series
in order to identify nonlinear dynamics of different brain states.
Liu et al. \cite{liu2011novel} proposed a pattern-based strategy
for prediction of short-duration and long-term activities in workflow
systems. Berndt \cite{berndt1994using} et al. used dynamic time warping
to identify patterns in time series. Fu et al. \cite{fu2007stock}
developed a framework for identification of the perceptually important
points in order to find patterns in stock market data and analyze
the trends. Alvisi et al. \cite{alvisi2007short} proposed a water-demand
forecasting model based on the periodic patterns observed at various
levels (annual, weekly and daily) and used it in a feed-forward control
system for decision-making. Alvarez et al. \cite{alvarez2010energy}
introduced a pattern-based approach to forecasting in time series
using the similarity of patterns to historical data. Hu et al.
\cite{hu2014pattern} used generalized principal component analysis
to identify the patterns in the historical wind speed data and used
them in the wind speed prediction. Among the aforementioned approaches, \cite{alvisi2007short,alvarez2010energy,hu2014pattern} are the only works that focused on using the observed patterns in data to forecast. However, their approaches require pattern classification methods and historical data as a reference. In this paper, we intend to use a binary structure tree to naturally classify and learn the patterns in the time series and use it for \textit{online} estimation without historical data requirement.

In this paper, we first introduce an analytic method to identify and
learn patterns in time series using a binary-structured tree (similar
to context tree \cite{WillemsAl97,WillemsAl95,Willems98}), and then
we present how these patterns can be used for estimation in time series
data. While assuming a single generative distribution with fixed parameters
for time series samples does not seem to be fully practical, it serve as our starting
point to see how such a hypothetical distribution may be decomposed
based on patterns in the time series. Albeit the focus of this paper
is on how patterns in the time series can be used for estimation,
the proposed pattern-based analysis can be used for many purposes
including compression and anomaly detection. In fact, our proposed methodology allows the extension of Context-Tree Weighting \cite{WillemsAl97,WillemsAl95,Willems98}
to real-valued case.

This paper is organized as follows: in Section \ref{sec:Notation},
the notation used in this paper is introduced. Section
\ref{sec:Thoery} develops a pattern-based
decomposition of distributions. Section \ref{sec:Estimation} represents
our proposed estimation method using the patterns in time series.
Finally, section \ref{sec:Experiment} presents numerical results.
\vspace{-0.02in}

\section{\label{sec:Notation}Notation}
\vspace{-0.02in}

For any binary variable $b\in\left\{ 0,1\right\} $, we define
\begin{align*}
x\overset{b}{\lessgtr}y & \triangleq\begin{cases}
x>y, & \qquad b=1;\\
x\leq y, & \qquad b=0.
\end{cases}\\
x\overset{b}{\lessgtr}y & \Longleftrightarrow y\overset{1-b}{\lessgtr}x.\\
x\overset{b}{\lessgtr}y & \Longleftrightarrow x\overset{b}{\gtrless}y.
\end{align*}
which characterizes the non-symmetric binary operators $\lessgtr$ and
$\gtrless$; note that $x\lessgtr y$ and $y\lessgtr x$ are different.
Suppose $B^{\left(D\right)}=b_{1}b_{2}\cdots b_{D}$ is a binary string
of length $D$, and let $\mathcal{B}^{\left(D\right)}$ be the set
of all possible binary strings of length $D$. For instance, $\mathcal{B}^{\left(2\right)}=\left\{ 00,01,10,11\right\} $.
In this paper, we use $B^{\left(D\right)}$ to indicate a pattern and
we consider two types of patterns: static pattern and dynamic pattern.
These patterns are defined as
\vspace{-0.1in}

\begin{align*}
B_{st}^{\left(D\right)}\left(i\right) & \triangleq\bigcap_{n=i-D+1}^{i}\left\{x_{i}\overset{b_{-n+i+1}}{\lessgtr}x_{n-1}\right\},\\
B_{dy}^{\left(D\right)}\left(i\right) & \triangleq\bigcap_{n=i-D+1}^{i}\left\{x_{n}\overset{b_{-n+i+1}}{\lessgtr}x_{n-1}\right\}.
\end{align*}
It follows that $B_{st}^{\left(D\right)}\left(i\right)=\bigcap\left\{x_{i}\overset{b_{1}}{\lessgtr}x_{i-1},B_{st}^{\left(D\right)}\left(i-1\right)\right\}$
and similarly $B_{dy}^{\left(D\right)}\left(i\right)=\bigcap\left\{x_{i}\overset{b_{1}}{\lessgtr}x_{i-1},B_{dy}^{\left(D\right)}\left(i-1\right)\right\}$.
As easily seen, at any time $i$, $B_{st}^{\left(D\right)}\left(i\right)$
is used to compare the current sample $x_{i}$ with previous $D$
samples, and $B_{dy}^{\left(D\right)}\left(i\right)$ is used to compare
the past $D$ consecutive pairs. While the dynamic pattern $B_{dy}^{\left(D\right)}\left(i\right)$, represents the \emph{actual} pattern in
the time series, we show below that it is closely tied to the static
pattern $B_{st}^{\left(D\right)}\left(i\right)$. Given depth $D$
and pattern $B^{\left(D\right)}$, we define the following probability
distribution functions (p.d.f.)
\begin{align*}
f^{\left(B_{st}^{\left(D\right)}\right)}\left(x\right) & =f\left( x=x_{i}\biggl|B_{st}^{\left(D\right)}\left(i\right)\right) \\
 & =f\left( x=x_{i}\biggl|\bigcap_{n=i-D+1}^{i}\left\{x_{i}\overset{b_{-n+i+1}}{\lessgtr}x_{n-1}\right\}\right), \\
f^{\left(B_{dy}^{\left(D\right)}\right)}\left(x\right) & =f\left( x=x_{i}\biggl|B_{dy}^{\left(D\right)}\left(i\right)\right) \\
 & =f\left( x=x_{i}\biggl|\bigcap_{n=i-D+1}^{i}\left\{x_{n}\overset{b_{-n+i+1}}{\lessgtr}x_{n-1}\right\}\right).
\end{align*}
Also, $f^{\left(B_{st}^{\left(D\right)}\right)}\left(x\right)$ and
$f_{st}^{\left(B^{\left(D\right)}\right)}\left(x\right)$ can be used
interchangeably (similarly $f^{\left(B_{dy}^{\left(D\right)}\right)}\left(x\right)$
might be used interchangeably with $f_{dy}^{\left(B^{\left(D\right)}\right)}\left(x\right)$).

\section{\label{sec:Thoery}Pattern-Based Decomposition of Distributions}
\vspace{-0.02in}

\begin{figure}[tbh]
\vspace{-0.15in}

\begin{centering}
\includegraphics[width=3.5in]{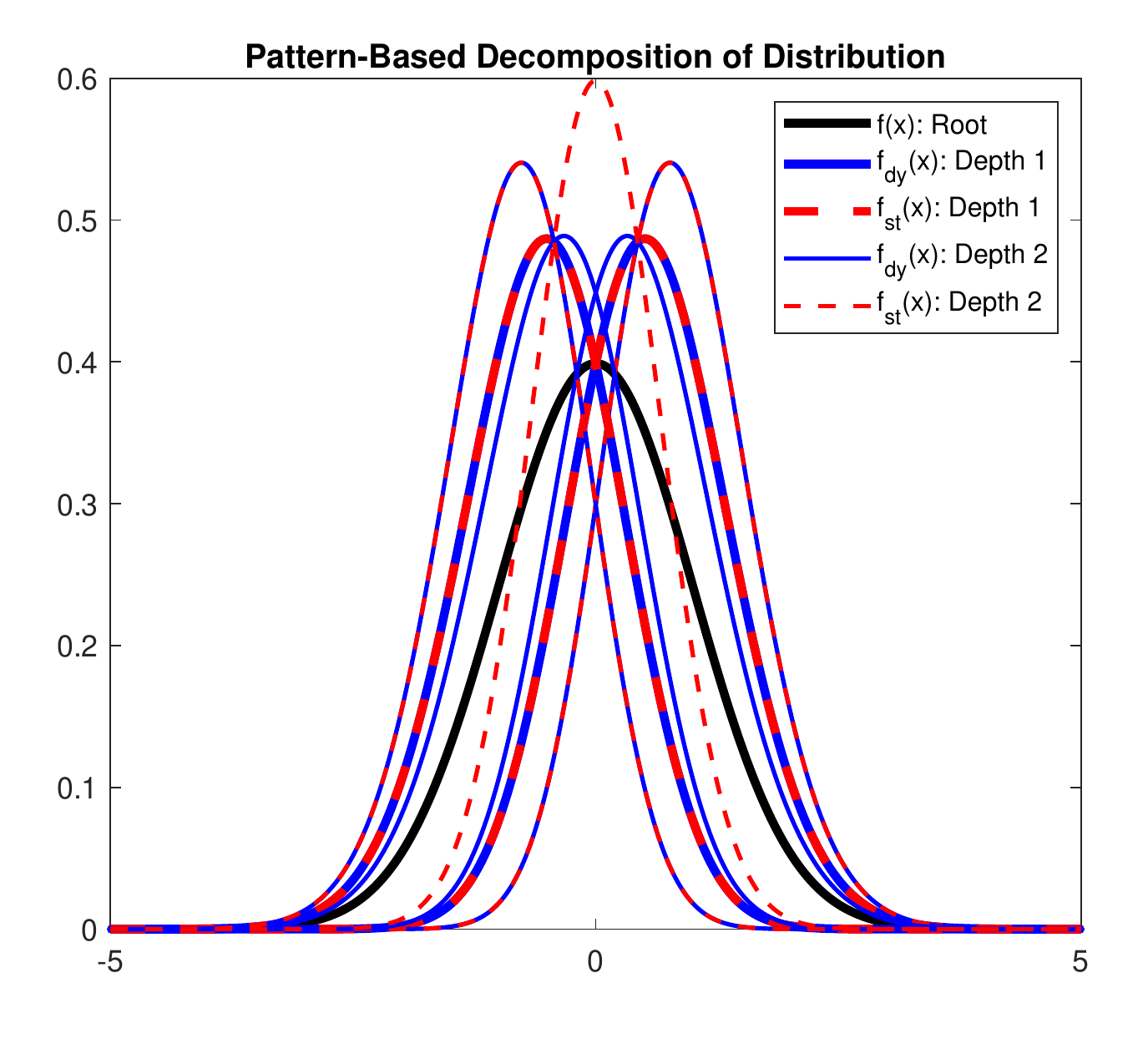}\vspace{-0.15in}
\par\end{centering}
\caption{\label{fig:DecompPDFs}p.d.f.s of the pattern-based decomposed samples
of standard normal time series (for both static and dynamic patterns at depth one and depth two):
$f\left(x\right)$, $f^{\left(0\right)}\left(x\right)$, $f^{\left(1\right)}\left(x\right)$,
$f^{\left(11\right)}\left(x\right)$, $f^{\left(10\right)}\left(x\right)$,
$f^{\left(01\right)}\left(x\right)$ and $f^{\left(00\right)}\left(x\right)$.}
\vspace{-0.1in}
\end{figure}

Suppose $x_i$, $i=1,2,\ldots, $ are time series samples drawn i.i.d.
according to the Gaussian distribution $\mathcal{N}\left(\mu,\sigma^{2}\right)$ with known mean
$\mu$ and variance $\sigma^{2}$ (the i.i.d. assumption will be relaxed
in Section \ref{sec:Estimation}). We call such a time series an independent \emph{normally
distributed time series}. Given any static or dynamic pattern $B^{\left(D\right)}$,
we would like to calculate the probability $\Pr\left( B^{\left(D\right)}\right) $ of observing such a pattern and the distribution $f^{\left(B^{\left(D\right)}\right)}\left(x\right)$
of samples compatible with such a pattern.
As presented in Theorem \ref{thm:static} and Theorem \ref{thm:dynamic},
these quantities depend on the beta-normal distribution \cite{eugene2002beta,gupta2005moments}
whose p.d.f. is given by
\vspace{-0.06in}
\begin{align}
\textrm{BN}\left(\alpha,\beta,\mu,\sigma\right) & \triangleq\frac{1}{\textrm{B}\left(\alpha,\beta\right)}\left[\Phi\left(\frac{x-\mu}{\sigma}\right)\right]^{\alpha-1}\nonumber \\
 & \times\left[1-\Phi\left(\frac{x-\mu}{\sigma}\right)\right]^{\beta-1}\frac{1}{\sigma}\phi\left(\frac{x-\mu}{\sigma}\right),\label{eq:BetaNornmal}
\end{align}
where $\phi\left(x\right)$ and $\Phi\left(x\right)$ are the standard
normal p.d.f. and c.d.f. respectively, and $\textrm{B}\left(\alpha,\beta\right)=\frac{\Gamma\left(\alpha\right)\Gamma\left(\beta\right)}{\Gamma\left(\alpha+\beta\right)}$
is the beta function. In Theorem \ref{thm:static}, given a static
pattern $B^{\left(D\right)}$, closed-form equations for $\Pr\left( B^{\left(D\right)}\right) $
and $f^{\left(B^{\left(D\right)}\right)}\left(x\right)$ are presented.
These results are then generalized to dynamic patterns in Theorem \ref{thm:dynamic}.
Due to the page limitation and similarity of the proof steps, we only provide the proof of Theorem \ref{thm:dynamic} which is more general than Theorem \ref{thm:static}.
\begin{thm}
\label{thm:static}Suppose $x_i$, $i=1,2,\ldots, $ are time series samples drawn i.i.d.
according to the Gaussian distribution $\mathcal{N}\left(\mu,\sigma^{2}\right)$. Given depth $D$ and any static pattern
$B_{st}^{\left(D\right)}$ we have
\begin{enumerate}
\item $\Pr\left( B_{st}^{\left(D\right)}\left(i\right)\right) =\textrm{B}\left(\alpha,\beta\right)$
\item $f^{\left(B_{st}^{\left(D\right)}\right)}\left(x\right)=\textrm{BN}\left(\alpha,\beta,\mu,\sigma\right)$
\end{enumerate}
where $\alpha=\sum_{i=1}^{D}b_{i}+1$ and $\beta=D-\sum_{i=1}^{D}b_{i}+1$
are integers.
\end{thm}
\begin{thm}
\label{thm:dynamic}Suppose $x_i$, $i=1,2,\ldots, $ are time series samples drawn i.i.d.
according to the Gaussian distribution $\mathcal{N}\left(\mu,\sigma^{2}\right)$. Given depth $D$ and any dynamic pattern
$B_{dy}^{\left(D\right)}$, the distribution $f^{\left(B_{dy}^{\left(D\right)}\right)}\left(x\right)$
can be written as the mixture of beta-normal distribution.
\end{thm}
\begin{IEEEproof}
In order to derive $f^{\left(B_{dy}^{\left(D\right)}\right)}\left(x\right)$,
we first need to find $\Pr\left( B_{dy}^{\left(D\right)}\left(i\right)\right) =\Pr\left( \bigcap_{n=i-D+1}^{i}\left\{x_{n}\overset{b_{-n+i+1}}{\lessgtr}x_{n-1}\right\}\right) $
(to reduce clutter, the definite integrals  $\int_{-\infty}^{+\infty}$ is abbreviated to $\int$){\footnotesize{}
\vspace{-0.12in}
\begin{align*}
A & = \Pr\left( B_{dy}^{\left(D\right)}\left(i\right)\right) \\
 & =\int\Pr\left( x_{i},B_{dy}^{\left(D\right)}\left(i\right)\right) dx_{i}\\
 & =\int\Pr\left( x_{i},x_{i}\overset{b_{1}}{\lessgtr}x_{i-1},B_{dy}^{\left(D\right)}\left(i-1\right)\right) dx_{i}\\
 & =\int\Pr\left( B_{dy}^{\left(D\right)}\left(i-1\right)\biggl|x_{i},x_{i}\overset{b_{1}}{\lessgtr}x_{i-1}\right) \\
 & \times\Pr\left( x_{i}\overset{b_{1}}{\lessgtr}x_{i-1}\biggl|x_{i}\right) f\left( x_{i}\right) dx_{i}\\
 & =\int\sum_{K}\Biggl[\Pr\left( B_{dy}^{\left(D\right)}\left(i-1\right)\biggl|x_{i},x_{i}\overset{b_{1}}{\lessgtr}x_{i-1},K_{st}^{\left(D-1\right)}\left(i-1\right)\right) \\
 & \times\Pr\left( K_{st}^{\left(D-1\right)}\left(i-1\right)\biggl|x_{i}\right) \Biggr]\Pr\left( x_{i}\overset{b_{1}}{\lessgtr}x_{i-1}|x_{i}\right) f\left( x_{i}\right) dx_{i}
\end{align*}
}where $\sum_{K}${\footnotesize{} }is an abbreviation for $\sum_{K^{\left(D-1\right)}\in\mathcal{B}^{\left(D-1\right)}}$, and $K_{st}^{\left(D-1\right)}\left(i-1\right)=\bigcap_{n=i-D+1}^{i-1}\left\{x_{i}\overset{k_{-n+i}}{\lessgtr}x_{n-1}\right\}$.
Therefore{\footnotesize{}
\begin{align*}
A & =\int\sum_{K}\Biggl[\Pr\left( B_{dy}^{\left(D\right)}\left(i-1\right)\biggl|x_{i},x_{i}\overset{b_{1}}{\lessgtr}x_{i-1},K_{st}^{\left(D-1\right)}\left(i-1\right)\right) \\
 & \times\Pr\left( K_{st}^{\left(D-1\right)}\left(i-1\right)\biggl|x_{i}\right) \Biggr]\Pr\left( x_{i}\overset{b_{1}}{\lessgtr}x_{i-1}|x_{i}\right) f\left( x_{i}\right) dx_{i}\\
 & =\int\sum_{K}\Biggl[\Pr\left( B_{dy}^{\left(D\right)}\left(i-1\right)\biggl|x_{i},x_{i}\overset{b_{1}}{\lessgtr}x_{i-1},K_{st}^{\left(D-1\right)}\left(i-1\right)\right) \\
 & \times\prod_{n=i-D+1}^{i-1}\Pr\left( x_{i}\!\!\!\overset{k_{-n+i}}{\lessgtr}\!\!\!x_{n-1}\biggl|x_{i}\right) \Biggr]\Pr\left( x_{i}\overset{b_{1}}{\lessgtr}x_{i-1}\biggl|x_{i}\right) f\left( x_{i}\right) dx_{i}\\
 & =\int\sum_{K}\Biggl[\Pr\left( x_{i-1}\overset{b_{2}}{\lessgtr}x_{i-2}\biggl|x_{i},x_{i}\overset{b_{1}}{\lessgtr}x_{i-1},x_{i}\overset{k_{1}}{\lessgtr}x_{i-2}\right) \\
 & \times\prod_{n=i-D+1}^{i-2}\Pr\left( x_{n}\!\!\!\!\overset{b_{-n+i+1}}{\lessgtr}\!\!\!\!x_{n-1}\biggl|x_{i},x_{i}\!\!\!\!\overset{k_{-n+i-1}}{\lessgtr}\!\!\!\!x_{n},x_{i}\!\!\!\!\overset{k_{-n+i}}{\lessgtr}\!\!\!\!x_{n-1}\right) \\
 & \times\prod_{n=i-D+1}^{i-1}\Pr\left( x_{i}\!\!\!\!\overset{k_{-n+i}}{\lessgtr}\!\!\!\!x_{n-1}\biggl|x_{i}\right) \Biggr]\Pr\left( x_{i}\overset{b_{1}}{\lessgtr}x_{i-1}\biggl|x_{i}\right) f\left( x_{i}\right) dx_{i}
\end{align*}
}where $\Pr\left( x_{i-1}\overset{b_{2}}{\lessgtr}x_{i-2}\biggl|x_{i},x_{i}\overset{b_{1}}{\lessgtr}x_{i-1},x_{i}\overset{k_{1}}{\lessgtr}x_{i-2}\right) $
and $\Pr\left( x_{n}\overset{b_{-n+i+1}}{\lessgtr}x_{n-1}\biggl|x_{i},x_{i}\overset{k_{-n+i-1}}{\lessgtr}x_{n},x_{i}\overset{k_{-n+i}}{\lessgtr}x_{n-1}\right) $
are either zero, or one or $\frac{1}{2}$ since for i.i.d. samples $\Pr\left( x_{l}>x_{m}\right) =\frac{1}{2}$
for any integer $l,m$. Thus we have 
\begin{align*}
\Pr\left( x_{n}>x_{n-1}|x_{i},x_{i}>x_{n},x_{i}>x_{n-1}\right)  & =\frac{1}{2};\\
\Pr\left( x_{n}>x_{n-1}|x_{i},x_{i}>x_{n},x_{i}<x_{n-1}\right)  & =0;\\
\Pr\left( x_{n}>x_{n-1}|x_{i},x_{i}<x_{n},x_{i}>x_{n-1}\right)  & =1;\\
\Pr\left( x_{n}>x_{n-1}|x_{i},x_{i}<x_{n},x_{i}<x_{n-1}\right)  & =\frac{1}{2};\\
\Pr\left( x_{n}<x_{n-1}|x_{i},x_{i}>x_{n},x_{i}>x_{n-1}\right)  & =\frac{1}{2};\\
\Pr\left( x_{n}<x_{n-1}|x_{i},x_{i}>x_{n},x_{i}<x_{n-1}\right)  & =1;\\
\Pr\left( x_{n}<x_{n-1}|x_{i},x_{i}<x_{n},x_{i}>x_{n-1}\right)  & =0;\\
\Pr\left( x_{n}<x_{n-1}|x_{i},x_{i}<x_{n},x_{i}<x_{n-1}\right)  & =\frac{1}{2}.
\end{align*}
Defining $\Psi\left(B^{\left(D\right)},K^{\left(D-1\right)}\right)\triangleq\Pr\left( x_{i-1}\overset{b_{2}}{\lessgtr}x_{i-2}\biggl|x_{i},x_{i}\overset{b_{1}}{\lessgtr}x_{i-1},x_{i}\overset{k_{1}}{\lessgtr}x_{i-2}\right) \times\prod_{n=i-D+1}^{i-2}\Pr\left( x_{n}\!\!\!\!\overset{b_{-n+i+1}}{\lessgtr}\!\!\!\!x_{n-1}\biggl|x_{i},x_{i}\!\!\!\!\overset{k_{-n+i-1}}{\lessgtr}\!\!\!\!x_{n},x_{i}\!\!\!\!\overset{k_{-n+i}}{\lessgtr}\!\!\!\!x_{n-1}\right) $,
we obtain{\footnotesize{}
\begin{align*}
A & =\sum_{K}\Psi\left(B^{\left(D\right)},K^{\left(D-1\right)}\right)\int\prod_{n=i-D+1}^{i-1}\Pr\left( x_{i}\overset{k_{-n+i}}{\lessgtr}x_{n-1}\biggl|x_{i}\right) \\
 & \times\Pr\left( x_{i}\overset{b_{1}}{\lessgtr}x_{i-1}\biggl|x_{i}\right) f\left( x_{i}\right) dx_{i}\\
 & =\sum_{K}\Psi\left(B^{\left(D\right)},K^{\left(D-1\right)}\right)\int\Phi\left(\frac{x_{i}-\mu}{\sigma}\right)^{\sum k_{j}+b_{1}}\\
 & \times\left[1-\Phi\left(\frac{x_{i}-\mu}{\sigma}\right)\right]^{\left(D-1-\sum k_{j}\right)+\left(1-b_{1}\right)}\frac{1}{\sigma}\phi\left(\frac{x_{i}-\mu}{\sigma}\right)dx_{i}\\
 & =\sum_{K^{\left(D-1\right)}\in\mathcal{B}^{\left(D-1\right)}}\Psi\left(B^{\left(D\right)},K^{\left(D-1\right)}\right)\textrm{B}\left(\alpha_{K},\beta_{K}\right),
\end{align*}
}where $\alpha_{K}=\sum_{j=1}^{D-1}k_{j}+b_{1}+1$ and $\beta_{K}=D-b_{1}-\sum_{j=1}^{D-1}k_{j}+1$
and $\textrm{B}\left(\alpha_{K},\beta_{K}\right)$ is the beta function.
Finally we need to calculate $\Psi\left(B^{\left(D\right)},K^{\left(D-1\right)}\right)$.
After some manipulations, $\Psi\left(B^{\left(D\right)},K^{\left(D-1\right)}\right)$
can be calculated recursively for $D\geq3$ as follows {\footnotesize{}
\begin{align}
\Psi\left(B^{\left(D\right)},K^{\left(D-1\right)}\right) & =\Psi\left(B^{\left(D-1\right)},K^{\left(D-2\right)}\right)\Biggl[\frac{1}{2}\left(k_{D-2}\odot k_{D-1}\right)\nonumber \\
 & +\left(k_{D-2}\oplus k_{D-1}\right)\left(k_{D-1}b_{D}+k_{D-2}\left(1-b_{D}\right)\right)\Biggr]\label{eq:Recursive}
\end{align}
}with $\Psi\left(11,1\right)=\frac{1}{2}$, $\Psi\left(11,0\right)=0$,
$\Psi\left(10,1\right)=\frac{1}{2}$ and $\Psi\left(10,0\right)=1$
initialization for $D<3$. Here $\oplus$ represents the xor operator
and $\odot$ represents the xnor operator. Thus, the distribution $f^{\left(B_{dy}^{\left(D\right)}\right)}\left(x\right)$
can be written as
\begin{align*}
f^{\left(B_{dy}^{\left(D\right)}\right)}\left(x\right) & =f\left( x=x_{i}\biggl|B_{dy}^{\left(D\right)}\right) \\
 & =\frac{\Pr\left( x_{i},B_{dy}^{\left(D\right)}\right) }{\Pr\left( B_{dy}^{\left(D\right)}\right) }\\
 & =\!\!\!\!\!\!\!\!\!\!\sum_{K^{\left(D-1\right)}\in\mathcal{B}^{\left(D-1\right)}}\!\!\!\!\!\!\!\!\!\!\Psi_{w}\left(B^{\left(D\right)},K^{\left(D-1\right)}\right)\textrm{BN}\left(\alpha_{K},\beta_{K},\mu,\sigma\right),
\end{align*}
where $\textrm{BN}\left(\alpha_{K},\beta_{K},\mu,\sigma\right)$ is
the beta-normal distribution and
\begin{align}
\Psi_{w}\left(B^{\left(D\right)},K^{\left(D-1\right)}\right) & =\frac{\Psi\left(B^{\left(D\right)},K^{\left(D-1\right)}\right)\textrm{B}\left(\alpha_{k},\beta_{k}\right)}{\sum_{L}\Psi\left(B^{\left(D\right)},L^{\left(D-1\right)}\right)\textrm{B}\left(\alpha_{L},\beta_{L}\right)}.\label{eq:Weighted}
\end{align}
Since $\sum_{K^{\left(D-1\right)}\in\mathcal{B}^{\left(D-1\right)}}\Psi_{w}\left(B^{\left(D\right)},K^{\left(D-1\right)}\right)=1$
the proof is complete. 
\end{IEEEproof}
\begin{example}
Assume $x_i$, $i=1,2,\ldots$ satisfies the assumptions of Theorem 2, and consider $f_{dy}^{\left(101\right)}\left(x\right)$.
Using Equation \eqref{eq:Recursive} we have $\Psi\left(101,11\right)=\Psi\left(10,1\right)\times\frac{1}{2}=\frac{1}{4}$,
$\Psi\left(101,10\right)=\Psi\left(10,1\right)\times0=0$, $\Psi\left(101,01\right)=\Psi\left(10,0\right)\times1=1$
and $\Psi\left(101,00\right)=\Psi\left(10,0\right)\times\frac{1}{2}=\frac{1}{2}$.
Hence using Equation \eqref{eq:Weighted} we can write
\begin{align*}
\Psi_{w}\left(101,11\right) & =\frac{\frac{1}{4}\textrm{B}\left(4,1\right)}{\frac{1}{4}\textrm{B}\left(4,1\right)+\textrm{B}\left(3,2\right)+\frac{1}{2}\textrm{B}\left(2,3\right)}=\frac{3}{9},\\
\Psi_{w}\left(101,10\right) & =0,\\
\Psi_{w}\left(101,01\right) & =\frac{\textrm{B}\left(3,2\right)}{\frac{1}{4}\textrm{B}\left(4,1\right)+\textrm{B}\left(3,2\right)+\frac{1}{2}\textrm{B}\left(2,3\right)}=\frac{4}{9},\\
\Psi_{w}\left(101,00\right) & =\frac{\frac{1}{2}\textrm{B}\left(2,3\right)}{\frac{1}{4}\textrm{B}\left(4,1\right)+\textrm{B}\left(3,2\right)+\frac{1}{2}\textrm{B}\left(2,3\right)}=\frac{2}{9}.
\end{align*}
Therefore the distribution $f_{dy}^{\left(101\right)}\left(x\right)$
is
\begin{align*}
f^{\left(101\right)}\left(x\right) & =\frac{3}{9}\textrm{BN}\left(4,1,\mu,\sigma\right)+\frac{4}{9}\textrm{BN}\left(3,2,\mu,\sigma\right)\\
 & +\frac{2}{9}\textrm{BN}\left(2,3,\mu,\sigma\right).
\end{align*}
Also note that for static pattern $f_{st}^{\left(101\right)}\left(x\right)=\textrm{BN}\left(3.2,\mu,\sigma\right)$.
\end{example}
Fig. \ref{fig:DecompPDFs} represents the p.d.f.s of the samples following the static and dynamic
depth-one and depth-two patterns. Obviously, at depth one the static and dynamic patterns are the same, and for $D\geq2$, only all-zero and all-one static and dynamic patterns result in the same distributions.
\begin{figure}[tbh]
\vspace{-0.15in}

\vspace{-0.1in}
\end{figure}

\section{\label{sec:Estimation}Pattern-Based Estimation Model}

So far we have assumed i.i.d. samples, in which for any $m\in\mathbb{N}$
such that $m<i$ we have $\Pr\left( x_{i}<x_{i-m}\right) =\Pr\left( x_{i}>x_{i-m}\right) =\frac{1}{2}$.
This leads to a symmetric decomposition of the distributions of interest. Note that
in many real-world applications such symmetric and i.i.d. assumptions
are not valid, especially in finite time series. In fact, such asymmetry
may be learned and used for prediction, estimation and compression.
In this paper, we only focus on estimation. From now on, by ``pattern'' we only mean dynamic pattern.

Consider the autoregressive time series $x_i$, $i=1,2,\ldots$, satisfying  $x_{i+1}=x_{i}+d_{i+1}$ where $d_{i}\sim\mathcal{N}\left(0,\sigma^{2}\right)$
is assumed to be i.i.d. and $x_{0}=\mu$. The $x_{i}$s are dependent
but identically distributed (d.i.d.), and given $x_{i}$ we have $x_{i+1}\sim\mathcal{N}\left(x_{i},\sigma^{2}\right)$. Due to the asymmetric patterns in time series,
instead of the model $x_{i+1}=x_{i}+d_{i+1}$ we use $x_{i+1}=x_{i}+\left(-1\right)^{q_{i+1}}\left|d_{i+1}\right|$
where $q_{i}\in\left\{ 0,1\right\} $ and $d_{i}\sim\mathcal{N}\left(0,\sigma^{2}\right)$.
Considering depth $D$ and given a pattern $B^{\left(D-1\right)}=b_{1}b_{2}\cdots b_{D-1}$,
the estimation (or \textit{forecast} as considered in \cite{alvisi2007short,alvarez2010energy}) of the next sample is $\widehat{x_{i+1}}=x_{i}+\left(-1\right)^{\widehat{q_{i+1}}}\left|\widehat{d_{i+1}}\right|$
where 
\begin{align*}
\widehat{q_{i+1}} & =\begin{cases}
0, & p_{i+1}^{\left(1\right)}>p_{i+1}^{\left(0\right)};\\
1, & \textrm{o.w.};
\end{cases}\\
p_{i+1}^{\left(1\right)} & =\Pr\left( x_{i+1}>x_{i}\biggl|B_{dy}^{\left(D-1\right)}\left(i\right)\right), \\
p_{i+1}^{\left(0\right)} & =1-p_{i+1}^{\left(1\right)},
\end{align*}
where $B_{dy}^{\left(D-1\right)}\left(i\right)=\bigcap_{n=i-D+2}^{i}\left\{x_{n}\overset{b_{-n+i+1}}{\lessgtr}x_{n-1}\right\}$
and
\begin{align*}
\widehat{d_{i+1}} & =\begin{cases}
\frac{1}{\left|S_{i+1}^{\left(1\right)}\right|}\sum_{x_{k}\in S_{i+1}^{\left(1\right)}}\left(x_{k}-x_{k-1}\right), & p_{i+1}^{\left(1\right)}>p_{i+1}^{\left(0\right)};\\
\frac{1}{\left|S_{i+1}^{\left(0\right)}\right|}\sum_{x_{k}\in S_{i+1}^{\left(0\right)}}\left(x_{k}-x_{k-1}\right), & \textrm{o.w.};
\end{cases}
\end{align*}
where $S_{i+1}^{\left(1\right)}=\left\{ x_{k}\biggl|1\leq k\leq i,\:x_{k}>x_{k-1},B_{dy}^{\left(D-1\right)}\left(k\right)\right\} $
and $S_{i+1}^{\left(0\right)}=\left\{ x_{k}\biggl|1\leq k\leq i,\:x_{k}\leq x_{k-1},B_{dy}^{\left(D-1\right)}\left(k\right)\right\} $.
Note that such an estimation comes down to two decisions: 1) Deciding
$x_{i}\overset{\widehat{q_{i+1}}}{\lessgtr}x_{i+1}$ by comparing $p_{i+1}^{\left(1\right)}$
and $p_{i+1}^{\left(0\right)}$ (where $p_{i+1}^{\left(1\right)}$
and $p_{i+1}^{\left(0\right)}$ are calculated using the results of Theorem \ref{thm:dynamic} for i.i.d. case, or calculated empirically for non-i.i.d. cases),
2) Calculating the change value $\widehat{d_{i+1}}$ by averaging
the change values for the samples that had the same pattern $\bigcap\left\{x_{i}\overset{\widehat{q_{i+1}}}{\lessgtr}x_{i+1},B_{dy}^{\left(D-1\right)}\left(i\right)\right\}$. 

\section{\label{sec:Experiment}Experiment}

In this section, we apply the proposed estimator 
on a synthetic time series, known as Mackey-Glass \cite{MackeyGlass},
as well as real-world time series of heart rate data. In these experiments,
first an empty binary-structured tree is created based on a (predetermined)
depth, then as samples of the time series are used iteratively to
fill out the tree based on the patterns, $p_{i+1}^{\left(1\right)}$ and $p_{i+1}^{\left(0\right)}$ are updated and the next sample is also
estimated using the samples and patterns that have already been seen,
and finally sample-wise estimation error is calculated. Algorithm \ref{alg:PT} summarized the step-by-step procedure. We compare the estimation results with linear prediction and an adapted version of the pattern-based forecasting method proposed in \cite{alvarez2010energy}. One of reasons for such an adaptation is that, the proposed method in \cite{alvarez2010energy} needs historical data, but in our online setting ``historical'' data becomes available iteratively as we see more data samples.

While it's not of our immediate interest and is subject of an ongoing
work, such a pattern tree can also be used in machine learning
framework, i.e. a training time series can be used to fill a \emph{pattern
tree}, and then the filled tree can be used in prediction/estimation
of a test time series. 

\begin{algorithm}
\caption{\label{alg:PT}Pattern Tree algorithms (Note that each pattern corresponds to a path in binary-structured tree).}
\begin{itemize}
\item \textbf{Inputs:}
\end{itemize}
$D_{max}$: The Predetermined maximum depth of tree \\
$x$: Time Series Data

\begin{algorithmic}[1]
\Procedure{Main}{} 
\State PT = Create($D_{max}$) 
\State PT = PT.SetPattern($x(1:D_{max})$)
\For{$i=D_{max}+1:End$}
\State PT = PT.Update($i,x(i)$)
\State $\widehat{x}(i+1)$ = PT.Estimate
\EndFor 
\EndProcedure
\end{algorithmic}
\vspace{.01in}

\begin{algorithmic}[1]
\Procedure{Create}{$D_{max}$} \Comment{Create an empty tree}
\State $n=2^{(D_{max}+1)}-1$ \Comment{Total number of nodes}
\State $I = 1:n$ \Comment{Index of nodes}
\State $Pattern = [\:]$ \Comment{Vector of size $D_{max}$}
\State $TS = [\:]$ \Comment{To record the time series}
\EndProcedure
\end{algorithmic}
\vspace{.01in}

\begin{algorithmic}[1]
\Procedure{SetPattern}{$x$} \Comment{Set the initial pattern}
\For{$i=2:D_{max}$}
\If{$x(i) > x(i-1)$} 
\State $Pattern(i-1) = 1$ 
\Else
\State $Pattern(i-1) = 0$
\EndIf  
\EndFor 
\State $TS = x$
\EndProcedure
\end{algorithmic}
\vspace{.01in}

\begin{algorithmic}[1]
\Procedure{Update}{$i,x$} 
\If{$x > TS(End)$} 
\State \textbf{update} $Pattern=[Pattern \quad 1]$ 
\Else
\State \textbf{update} $Pattern=[Pattern \quad 0]$
\EndIf 
\State \textbf{calculate} \textit{Path} using \textit{Pattern} 
\For{$d=1:D_{max}$}
\State \textbf{add} $i$ to the node $I(d,path)$
\EndFor
\State \textbf{update} $Pattern=Pattern(2:D_{max})$ 
\State \textbf{update} $TS = [TS \quad x]$
\EndProcedure
\end{algorithmic}
\vspace{.01in}

\begin{algorithmic}[1]
\Procedure{Estimate}{} 
\State $Pattern0 = [Pattern \quad 0]$
\State $Pattern1 = [Pattern \quad 1]$
\State \textbf{calculate} \textit{Path0} using \textit{Pattern0}
\State \textbf{calculate} \textit{Path1} using \textit{Pattern1}
\State $d = D_{max}$,   $J = [\:]$
\While{$d \geq 1$ and $J = [\:]$}
\If{$P(Pattern1)>P(Pattern0)$}
\State $J =$ indexes in node $I(d,path1)$
\State $\widehat{x} = TS(End) + mean(TS(J)-TS(J-1))$
\ElsIf{$P(Pattern0)>P(Pattern1)$}
\State $J =$ indexes in node $I(d,path0)$
\State $\widehat{x} = TS(End) - mean(TS(J)-TS(J-1))$
\EndIf
\State $d = d - 1$
\EndWhile
\State \textbf{return} $\widehat{x}$
\EndProcedure
\end{algorithmic}
\end{algorithm}

\subsection{\label{subsec:MGexperiment}Mackey-Glass}

The Mackey-Glass time series is a nonlinear time delay differential equation
and was originally introduced to represent the appearance of complex
dynamic in physiological control systems. It is derived by finite difference discretization of the nonlinear
differential equation $\frac{dx\left(t\right)}{dt}=-ax\left(t\right)+\frac{bx\left(t-\tau\right)}{1+x^{10}\left(t-\tau\right)},\:t\geq0$,
where $a$, $b$ and $\tau$ are constants. We generated 10000 samples
of this time series with $a=0.2$, $b=0.1$ and $\tau=17$ and followed
the steps described in Section \ref{sec:Experiment}. Table \ref{tab:Comparison}
summarizes the comparison between our proposed pattern tree method, the pattern-based forecasting method proposed in \cite{alvarez2010energy}
and linear prediction in terms of estimation mean squared error (MSE).
As can be seen, our proposed method outperforms others.
\begin{figure}[tbh]
\vspace{-0.15in}

\begin{centering}
\includegraphics[width=3.5in]{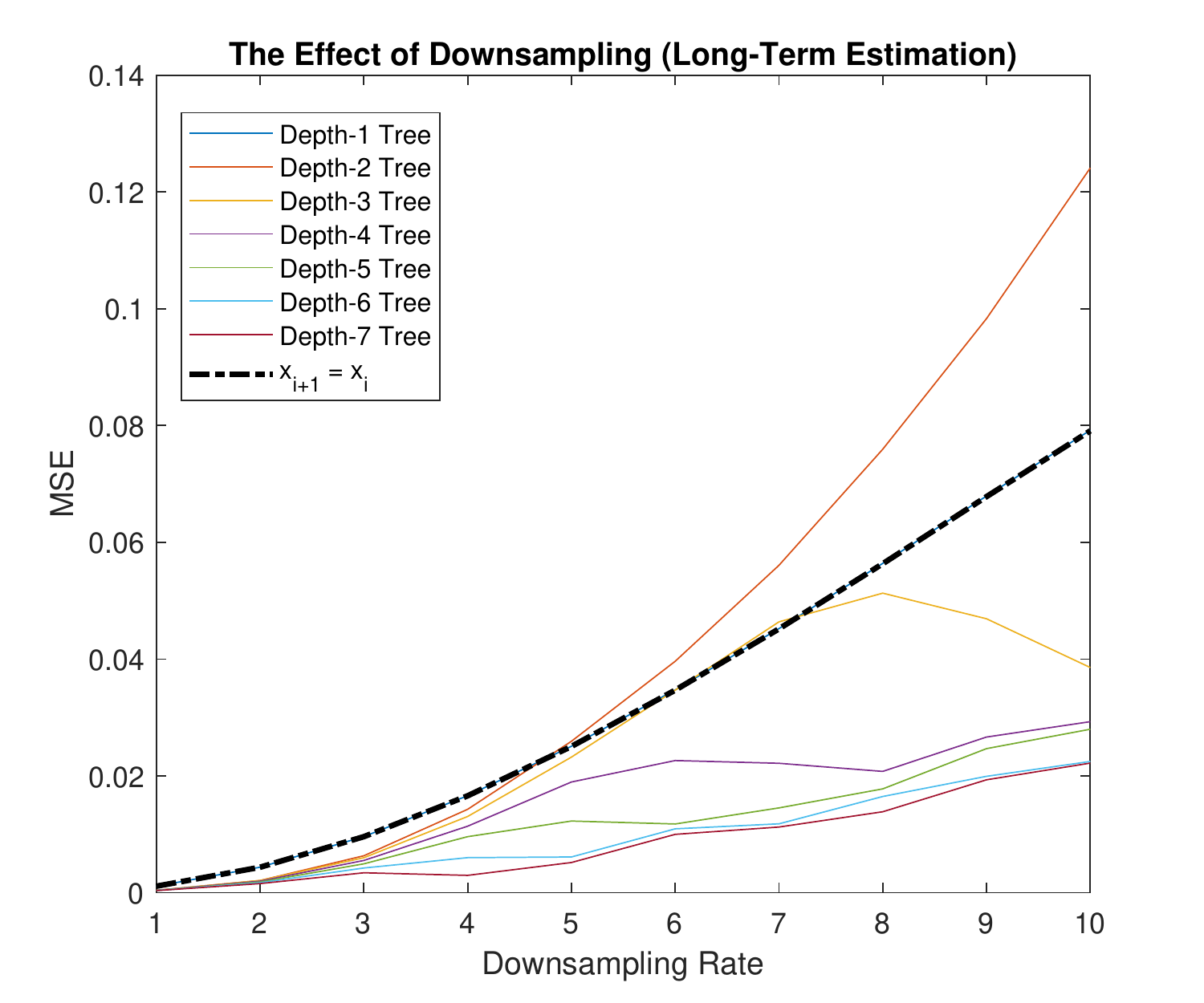}\vspace{-0.15in}
\par\end{centering}
\caption{\label{fig:MacketGlass}The effect of time series sampling frequency
on estimation error for pattern trees of various depths.}

\vspace{-0.1in}
\end{figure}

In the second experiment using Mackey-Glass time series, we analyze the effect of downsampling. Fig. \ref{fig:MacketGlass} shows the estimation MSE
versus downsampling rate. As expected, estimation using deeper pattern trees
($D\geq3$) are more resilient to downsampling.

\subsection{Heart Rate Data}

In the second experiment we used the heart rate time series
recorded by E4 Empatica wristbands (sampling frequency for hear rate
measurements of this wearable device is 1 Hz). Similar to the Mackey-Glass
experiment, we used 10000 samples of recorded heart rate data and followed
the steps described in Section \ref{sec:Experiment}. As can be seen in Table \ref{tab:Comparison}, similar to the previous experiment with Mackey-Glass time series, our proposed pattern tree
method performs better than others in terms of estimation MSE.

\begin{table}[tbh]
\begin{centering}
\begin{tabular}{|c|c|c|c|c|c|c|}
\hline 
\multirow{2}{*}{\thead{Depth/ \\ Order}} & \multicolumn{3}{c|}{Mackey-Glass} & \multicolumn{3}{c|}{Heart Rate}\tabularnewline
\cline{2-7} \cline{3-7} \cline{4-7} \cline{5-7} \cline{6-7} \cline{7-7} 
 & PT & \cite{alvarez2010energy} & LP & PT & \cite{alvarez2010energy} & LP\tabularnewline
\hline 
\hline
1 & $ \!\! 0.0011 \!\! $ & $ \!\! 0.0012 \!\! $ & $ \!\! 0.0011 \!\! $ & $ \!\! 0.0101 \!\! $ & $ \!\! 0.0114 \!\! $ & $ \!\! 0.2528 \!\! $\tabularnewline
\hline 
2 & $ \!\! 3.95 \!\! \times \!\! 10^{-4} \!\! $ & $ \!\! 6.15 \!\! \times \!\! 10^{-4} \!\! $ & $ \!\! 0.0081 \!\! $ & $ \!\! 0.0057 \!\! $ & $ \!\! 0.0104 \!\! $ & $ \!\! 0.3915 \!\! $\tabularnewline
\hline 
3 & $ \!\! 3.85 \!\! \times \!\! 10^{-4} \!\! $ & $ \!\! 6.17 \!\! \times \!\! 10^{-4} \!\! $ & $ \!\! 0.0153 \!\! $ & $ \!\! 0.0055 \!\! $ & $ \!\! 0.0103 \!\! $ & $ \!\! 0.3257 \!\! $\tabularnewline
\hline 
4 & $ \!\! 3.84 \!\! \times \!\! 10^{-4} \!\! $ & $ \!\! 6.11 \!\! \times \!\! 10^{-4} \!\! $ & $ \!\! 0.0239 \!\! $ & $ \!\! 0.0054 \!\! $ & $ \!\! 0.0103 \!\! $ & $ \!\! 0.3279 \!\! $\tabularnewline
\hline 
5 & $ \!\! 3.81 \!\! \times \!\! 10^{-4} \!\! $ & $ \!\! 5.95 \!\! \times \!\! 10^{-4} \!\! $ & $ \!\! 0.0336 \!\! $ & $ \!\! 0.0054 \!\! $ & $ \!\! 0.0102 \!\! $ & $ \!\! 0.3674 \!\! $\tabularnewline
\hline 
\end{tabular}
\par\end{centering}
\caption{\label{tab:Comparison}Comparison of mean squared error of estimation
using our proposed pattern tree (PT) method, linear prediction
(LP) and an adapted version of the pattern-based forecasting method proposed in \cite{alvarez2010energy} for various depths/orders.}
\end{table}


\IEEEtriggeratref{3}

\bibliographystyle{IEEEtran}
\bibliography{Prometheus,ECGandHRV,BigData}

\begin{thebibliography}{10}
\providecommand{\url}[1]{#1}
\csname url@samestyle\endcsname
\providecommand{\newblock}{\relax}
\providecommand{\bibinfo}[2]{#2}
\providecommand{\BIBentrySTDinterwordspacing}{\spaceskip=0pt\relax}
\providecommand{\BIBentryALTinterwordstretchfactor}{4}
\providecommand{\BIBentryALTinterwordspacing}{\spaceskip=\fontdimen2\font plus
\BIBentryALTinterwordstretchfactor\fontdimen3\font minus
  \fontdimen4\font\relax}
\providecommand{\BIBforeignlanguage}[2]{{%
\expandafter\ifx\csname l@#1\endcsname\relax
\typeout{** WARNING: IEEEtran.bst: No hyphenation pattern has been}%
\typeout{** loaded for the language `#1'. Using the pattern for}%
\typeout{** the default language instead.}%
\else
\language=\csname l@#1\endcsname
\fi
#2}}
\providecommand{\BIBdecl}{\relax}
\BIBdecl

\bibitem{sabeti2019data}
E.~Sabeti and A.~H{\o}st-Madsen, ``Data discovery and anomaly detection using
  atypicality for real-valued data,'' \emph{Entropy}, vol.~21, no.~3, p. 219,
  2019.

\bibitem{host2019data}
A.~H{\o}st-Madsen, E.~Sabeti, and C.~Walton, ``Data discovery and anomaly
  detection using atypicality: Theory,'' \emph{IEEE Transactions on Information
  Theory}, 2019.

\bibitem{islam2015internet}
S.~R. Islam, D.~Kwak, M.~H. Kabir, M.~Hossain, and K.-S. Kwak, ``The internet
  of things for health care: a comprehensive survey,'' \emph{IEEE Access},
  vol.~3, pp. 678--708, 2015.

\bibitem{breiman2017classification}
L.~Breiman, \emph{Classification and regression trees}.\hskip 1em plus 0.5em
  minus 0.4em\relax Routledge, 2017.

\bibitem{chung2004evolutionary}
F.-L. Chung, T.-C. Fu, V.~Ng, and R.~W. Luk, ``An evolutionary approach to
  pattern-based time series segmentation,'' \emph{IEEE transactions on
  evolutionary computation}, vol.~8, no.~5, pp. 471--489, 2004.

\bibitem{ouyang2010ordinal}
G.~Ouyang, C.~Dang, D.~A. Richards, and X.~Li, ``Ordinal pattern based
  similarity analysis for eeg recordings,'' \emph{Clinical Neurophysiology},
  vol. 121, no.~5, pp. 694--703, 2010.

\bibitem{liu2011novel}
X.~Liu, Z.~Ni, D.~Yuan, Y.~Jiang, Z.~Wu, J.~Chen, and Y.~Yang, ``A novel
  statistical time-series pattern based interval forecasting strategy for
  activity durations in workflow systems,'' \emph{Journal of Systems and
  Software}, vol.~84, no.~3, pp. 354--376, 2011.

\bibitem{berndt1994using}
D.~J. Berndt and J.~Clifford, ``Using dynamic time warping to find patterns in
  time series.'' in \emph{KDD workshop}, vol.~10, no.~16.\hskip 1em plus 0.5em
  minus 0.4em\relax Seattle, WA, 1994, pp. 359--370.

\bibitem{fu2007stock}
T.-c. Fu, F.-l. Chung, R.~Luk, and C.-m. Ng, ``Stock time series pattern
  matching: Template-based vs. rule-based approaches,'' \emph{Engineering
  Applications of Artificial Intelligence}, vol.~20, no.~3, pp. 347--364, 2007.

\bibitem{alvisi2007short}
S.~Alvisi, M.~Franchini, and A.~Marinelli, ``A short-term, pattern-based model
  for water-demand forecasting,'' \emph{Journal of hydroinformatics}, vol.~9,
  no.~1, pp. 39--50, 2007.

\bibitem{alvarez2010energy}
F.~M. Alvarez, A.~Troncoso, J.~C. Riquelme, and J.~S.~A. Ruiz, ``Energy time
  series forecasting based on pattern sequence similarity,'' \emph{IEEE
  Transactions on Knowledge and Data Engineering}, vol.~23, no.~8, pp.
  1230--1243, 2010.

\bibitem{teng2003regression}
W.-G. Teng, M.-S. Chen, and P.~S. Yu, ``A regression-based temporal pattern
  mining scheme for data streams,'' in \emph{Proceedings of the 29th
  international conference on Very large data bases-Volume 29}.\hskip 1em plus
  0.5em minus 0.4em\relax VLDB Endowment, 2003, pp. 93--104.

\bibitem{hu2014pattern}
Q.~Hu, P.~Su, D.~Yu, and J.~Liu, ``Pattern-based wind speed prediction based on
  generalized principal component analysis,'' \emph{IEEE Transactions on
  Sustainable Energy}, vol.~5, no.~3, pp. 866--874, 2014.

\bibitem{kozat2007universal}
S.~S. Kozat, A.~C. Singer, and G.~C. Zeitler, ``Universal piecewise linear
  prediction via context trees,'' \emph{IEEE Transactions on Signal
  Processing}, vol.~55, no.~7, pp. 3730--3745, 2007.

\bibitem{WillemsAl97}
F.~Willems, Y.~Shtarkov, and T.~Tjalkens, ``Reflections on "the context tree
  weighting method: Basic properties",'' \emph{Newsletter of the IEEE
  Information Theory Society}, vol.~47, no.~1, 1997.

\bibitem{WillemsAl95}
F.~M.~J. Willems, Y.~Shtarkov, and T.~Tjalkens, ``The context-tree weighting
  method: basic properties,'' \emph{Information Theory, IEEE Transactions on},
  vol.~41, no.~3, pp. 653--664, 1995.

\bibitem{Willems98}
F.~Willems, ``The context-tree weighting method: extensions,''
  \emph{Information Theory, IEEE Transactions on}, vol.~44, no.~2, pp.
  792--798, Mar 1998.

\bibitem{eugene2002beta}
N.~Eugene, C.~Lee, and F.~Famoye, ``Beta-normal distribution and its
  applications,'' \emph{Communications in Statistics-Theory and methods},
  vol.~31, no.~4, pp. 497--512, 2002.

\bibitem{gupta2005moments}
A.~K. Gupta and S.~Nadarajah, ``On the moments of the beta normal
  distribution,'' \emph{Communications in Statistics-Theory and Methods},
  vol.~33, no.~1, pp. 1--13, 2005.

\bibitem{MackeyGlass}
M.~C. Mackey and L.~Glass, ``Oscillation and chaos in physiological control
  systems,'' \emph{Science}, vol. 197, no. 4300, pp. 287--289, 1977.

\end{thebibliography}

\end{document}